# Carbon Monoxide Affecting Planetary Atmospheric Chemistry

Chao He[1], Sarah M. Hörst[1], Sydney Riemer[1], Joshua A. Sebree[2], Nicholas Pauley[2], Véronique Vuitton[3]

[1] Department of Earth and Planetary Sciences, Johns Hopkins University, Baltimore, MD, USA che13@jhu.edu

[2] Department of Chemistry and Biochemistry, University of Northern Iowa, Cedar Falls, IA, USA

[3] Université Grenoble Alpes, CNRS, IPAG, F-38000 Grenoble, France

**Abstract:**

CO is an important component in many $N_2/CH_4$ atmospheres including Titan, Triton, and Pluto, and has also been detected in the atmosphere of a number of exoplanets. Numerous experimental simulations have been carried out in the laboratory to understand the chemistry in $N_2/CH_4$ atmospheres, but very few simulations have included CO in the initial gas mixtures. The effect of CO on the chemistry occurring in these atmospheres is still poorly understood. We have investigated the effect of CO on both gas and solid phase chemistry in a series of planetary atmosphere simulation experiments using gas mixtures of CO, $CH_4$, and $N_2$ with a range of CO mixing ratios from 0.05% to 5% at low temperature (~100 K). We find that CO affects the gas phase chemistry, the density, and the composition of the solids. Specifically, with the increase of CO in the initial gases, there is less $H_2$ but more $H_2O$, HCN, $C_2H_5N$/HCNO and $CO_2$ produced in the gas phase, while the density, oxygen content, and degree of unsaturation of the solids increase. The results indicate that CO has an important impact on the chemistry occurring in our experiments and accordingly in planetary atmospheres.

1. INTRODUCTION

Atmospheric hazes are present in a range of solar system and exoplanetary atmospheres. The chemistry in planetary atmospheres induced by stellar radiation and energetic electrons produces the haze particles. Haze particles affect the physical and chemical processes occurring in the atmospheres and on the surface if planets are terrestrial. Organic hazes are particularly interesting due to astrobiological implications, such as the potential for Titan's atmosphere to contain the building blocks of life (Hörst et al. 2012).

The formation and the properties of haze particles in $N_2/CH_4$ atmospheres have been studied for decades in the laboratory through the production of Titan aerosol analogues or "tholins" (Cable et al. 2012), but these studies usually did not include other minor atmospheric constituents, such as carbon monoxide (CO). CO is an important component in the hazy, $N_2/CH_4$ atmospheres of Titan (~50 ppm) (Lutz et al. 1983; de Kok et al. 2007; Hörst 2017), Pluto (~0.05%) (Greaves et al. 2011; Lellouch et al. 2011, 2017), and Triton (~0.06%) (Lellouch et al. 2010). CO has also been detected in the atmosphere of a number of exoplanets (Madhusudhan & Seager 2009; Madhusudhan et al. 2011; de Kok et al. 2013; Konopacky et al. 2013; Travis & Barman 2015), and haze layers are suggested to be present in exoplanetary atmospheres (Pont et al. 2008, 2013; Deming et al. 2013). CO is of particular interest because it could serve as a source of oxygen for incorporation into haze particles and oxygen is an essential element of life in addition to carbon, hydrogen and nitrogen. Therefore, it is important to investigate the effect of CO on planetary atmospheric chemistry.

A few Titan atmosphere simulation experiments that include CO in the initial gases have been carried out. Bernard et al. (2003) and Coll et al. (2003) identified an O-containing compound, ethylene oxide ($C_2H_4O$), in the gas phase products of their $N_2$-$CH_4$-CO experiments. Tran et al. (2008) showed the formation of several aldehydes and ketones in the photolysis products of $N_2$-$CH_4$-$H_2$-$C_2H_2$-$C_2H_4$-$HC_3N$-CO. Hörst et al. (2012) detected amino acids and nucleotide bases in the solid products of their $N_2$-$CH_4$-CO experiments. Hörst and Tolbert (2014) focused on the effect of CO on the size and number density of haze particles. Fleury et al. (2014) focused on the influence of CO on the $N_2$-$CH_4$ reactivity. However, the effect of CO on the chemistry occurring in the system and the resulting gas and solid compositions are still not clear. Here we present a study of the effect of CO on the planetary atmospheric chemistry by measuring the composition of both gas and solid phase products in a series of atmosphere simulation experiments using AC glow discharge to initiate reactions in a range of $N_2$-$CH_4$-CO gas mixtures.

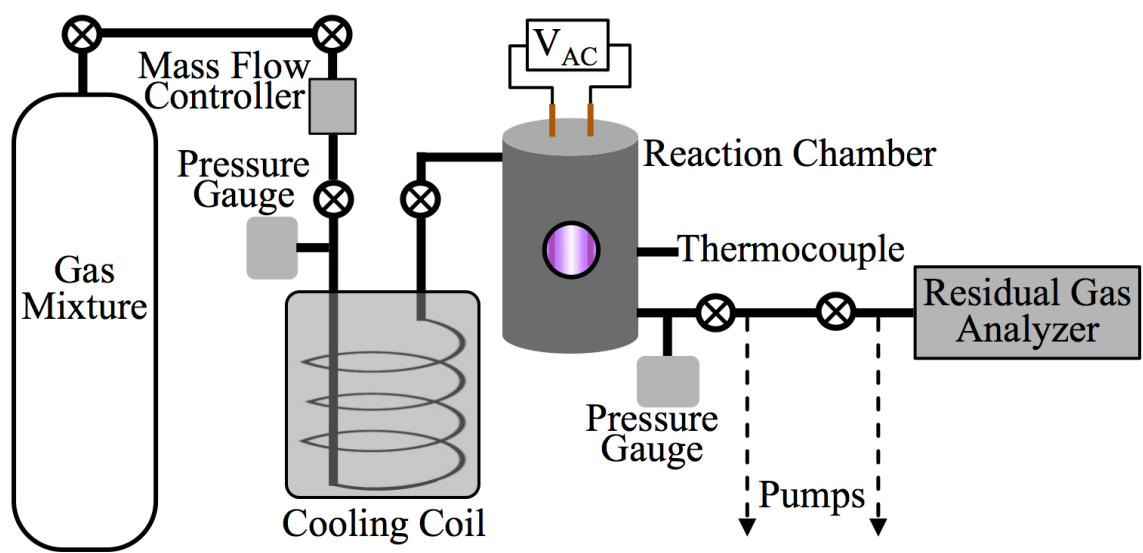

**Figure 1.** Schematic of the experimental setup used for this work. $N_2$, $CH_4$, and CO mix overnight in the mixing cylinder. Gases flow through a cooling coil immersed in liquid nitrogen (77 K) and into the reaction chamber where they are exposed to AC glow discharge initiating chemical processes that lead to the formation of new gas phase products and particles. The temperature of the gases in the reaction chamber was determined based on the pressure–temperature gas law ($P_1/P_2=T_1/T_2$) before starting experiment. After 1 hour flowing of the cold gases, we close the valves on both sides of the chamber and measure the pressure ($P_1$) of the cold gases in the chamber. We measure the pressure ($P_2$) in the chamber again when the gases are warmed up to 294 K ($T_2$, room temperature). The temperature ($T_1$) of the cold gases in the chamber is calculated to be 98±3 K. The gas phase products are monitored by a Residual Gas Analyzer (RGA, a mass spectrometer), and the solid particles deposited on the chamber wall are collected and stored in a dry $N_2$ atmosphere glove box for further analysis.

2. MATERIALS AND EXPERIMENTAL METHODS

*2.1. Haze Production Setup*

Figure 1 shows a schematic of Planetary HAZE Research (PHAZER) experimental setup, a new setup built at Johns Hopkins University. The system is constructed from stainless steel and the seals are copper gaskets, to avoid any contamination from plastic and maintain ultra high vacuum of the system. The gas mixture of CO, $CH_4$, and $N_2$ with a range of CO mixing ratios was prepared in a stainless steel cylinder. We introduce $CH_4$ (99.99% Airgas) with 5% mixing ratio and CO (99.99% Airgas) with a range of CO mixing ratios (from 0, 0.05%, 0.2%, 0.5%, 1%, 2.5%, to 5% in seven experiments) into the cylinder, and then fill the cylinder to 200 psi with the $N_2$ (99.999% Airgas). We allow the gases to mix for a minimum of twelve hours before running experiment. The gases continuously flow through a 15-meter stainless steel cooling coil immersed in liquid nitrogen (77 K), which cools the gases down to about 100 K and removes trace impurities in the gases. The cold gases then flow through a stainless steel reaction chamber where they are exposed to AC glow discharge initiating chemical processes that lead to the formation of new gas phase products and particles. The gas flow rate is maintained at 10 standard cubic centimeters per minute (sccm) by a mass flow controller (MKS Instrument, GM50A), so that the pressure in the reaction chamber is held at 2 Torr. This allows the reactant gases to be exposed to the glow discharge for about 3 seconds. After flowing through the chamber, most of the gases, including the newly formed gas products, are pumped out of the system by a dry scroll pump (Agilent, IDP-3). A small portion of the gases flows into a Residual Gas Analyzer (SRS, RGA300) where the gas phase products are monitored. RGA300 is a small quadrupole mass spectrometer with a standard 70 ev electron ionization source, covering 1–300 atomic mass unit (amu) with 0.5 amu resolution and $10^{-11}$ torr minimum detectable partial pressure. The RGA background mass spectrum (50-scan average) is recorded at a few $10^{-7}$ Torr. The gases flowing into RGA are controlled by a needle valve to maintain the pressure at a few $10^{-5}$ Torr in the RGA. The gas mixture mass spectrum (50-scan average) is collected before turning on the discharge. After turning on the discharge, we flow the gases for 30 minutes before we start the RGA scanning, and each scan takes about 2 minutes. In each experiment, an average mass spectrum (MS) is obtained after over 1000 scans.

After 72 hours of discharge flow, a red/brown film is deposited on the wall of the reaction chamber. The chamber is under vacuum for 48 hours to remove the volatile components, and then transferred to a glove box (Inert Technology Inc., I-lab 2GB) where the solids are collected under a dry $N_2$ atmosphere. The solids are weighed by an analytical balance (Sartorius Entris 224-1S with standard deviation of 0.1 mg), and kept in the glove box and wrapped in foil to avoid exposure to air and light, respectively.

*2.2. Density Measurements and Elemental Analysis of the Solids*

To get the density of each solid sample, we measure the mass and the volume of certain amount of sample by using the analytical balance and a gas pycnometer (AccPyc II 1340, Micrometrics), respectively. The gas pycnometer uses the gas (helium) displacement method to measure volume based on Boyle's law of volume–pressure relationships. We fill the sample to ~70% of a 0.1 cc cup, measure the mass of the sample in the cup on the balance (the mass of the cup is known), and measure the volume of each sample at ambient temperature by using the gas pycnometer. The volume is an average volume of 20 measurements with 0.00001 mL standard deviation. The density is equal to mass divided by volume.

The starting material is the gas mixture of CO, $CH_4$, and $N_2$, thus the solid material likely only contains C, H, N, and O atoms. We quantify the contents of four elements in each sample by using an organic elemental analyzer (Flash 2000 Elemental Analyzer, Thermo Scientific). The sample under test (~1 to 2 mg) is weighed in a tin capsule, and then the capsule enclosing the sample is placed in the analyzer for combustion analysis. We run 3 times for each sample. The analyzer gives the percentages of C, H, and N directly, and the percentage of O is determined by mass subtraction as the solids only contain C, H, N, and O. To minimize the adherence of water vapor, samples are kept under dry $N_2$ atmosphere until weighing.

*2.3. Nuclear Magnetic Resonance (NMR) Measurements of the Solids*

Approximately 20 mg of the solid sample is dissolved into 1 mL deuterated dimethyl sulfoxide (DMSO-$d_6$, D 99.9%, anhydrous, Sigma-Aldrich). The solid sample is totally dissolved and the solution is sealed in a NMR tube. The operation is performed in the glove box to prevent air exposure. For each sample, one-dimensional proton (1D $^1$H) and two-dimensional proton-carbon heteronuclear single quantum correlation (2D $^1$H–$^{13}$C HSQC where each signal represents a proton that is bound to a carbon atom) NMR measurements are performed on a Bruker Avance 600 MHz spectrometer with cryogenic probe with z-gradient. For the 1D $^1$H NMR spectra, the relaxation delay between each pulse is set to 30 seconds, which ensures that all types of protons relax completely before each pulse and the peaks in the spectra are quantitative. The chemical shift range is set to 0 to 12 ppm for $^1$H and 0 to 240 ppm for $^{13}$C, covering general organic species (Lambert & Mazzola 2004).

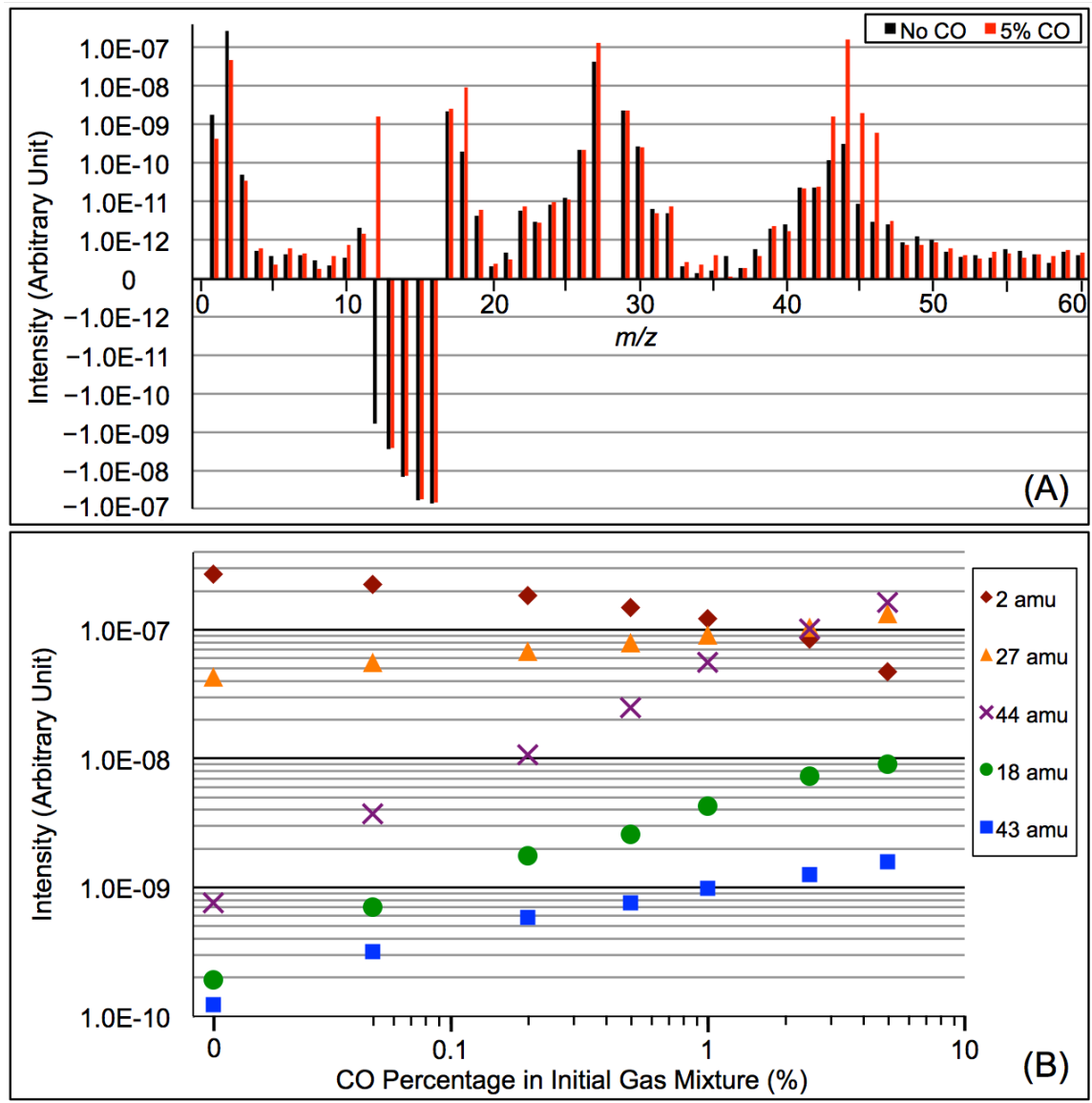

**Figure 2.** (A) The subtracted mass spectra (the mass spectrum with discharge off is subtracted from that with discharge on) for gas mixtures without CO (black) and with 5% CO (red). MS peaks from 1 to 60 amu are shown since the heavier peaks are near the noise level. Note: 28 amu is used as a fixed reference for normalization and therefore the intensity shown is 0. The peaks at 12-16 amu are from CH$_4$ and its fragments. In the subtracted spectra they are negative because their abundance decreases when the discharge turns on because some CH$_4$ molecules are dissociated and converted to other species. The peak at 12 amu in the spectrum with 5% CO is positive due to fragmentation of CO. (B) The peak intensity change with the increase of CO in the initial gases (The error bars in Figure 2B are about 0.8% to 1.7%, and are smaller than the size of the symbols). As CO in the initial gases increases, the peaks at 18, 27, 43 and 44 amu increase while the peak at 2 amu decreases. The peak at 2 amu corresponds to hydrogen molecule ($H_2$), 18 amu to water ($H_2O$), 27 amu to hydrogen cyanide (HCN), 43 amu to $C_2H_5N$ and/or HCNO and 44 amu to carbon dioxide ($CO_2$), respectively.

## 3. RESULTS AND DISCUSSION

### 3.1. Mass Spectra of Gas Phase Products

We recorded the mass spectra of the gas mixture with discharge off and on during each experiment. The RGA background is removed from both spectra and MS peaks in the spectra are normalized to the peak intensity at 28 amu. The peak at 28 amu is derived from $N_2$ and CO, which account for 95% in the initial gases. The intensity of the mass peak at 28 amu should therefore be constant enough for a fixed reference, as done also by Peng et al. (2013). The mass spectrum with discharge off is subtracted from that with discharge on and a subtracted mass spectrum is obtained for each gas mixture. Figure 2A shows the subtracted mass spectra for two gas mixtures, without CO and with 5% CO. Since we are trying to understand the effect of CO on the chemistry in the system, we focus on the peaks that have significant intensity change after introducing CO, such as the peaks at 1, 2, 12, 18, 27, 43, 44, 45 and 46 amu. The changed peaks correspond to the species in the gas phase, 2 amu to hydrogen molecule ($H_2$), 18 amu to water ($H_2O$), 27 amu to hydrogen cyanide (HCN), and 43 amu to $C_2H_5N$ and/or HCNO based on the nitrogen rule (an odd nominal mass indicates an odd number of nitrogen atoms are present), respectively. The peak at 44 amu could be from propane ($C_3H_8$), carbon dioxide ($CO_2$), acetaldehyde ($C_2H_4O$), and/or ethylene oxide ($C_2H_4O$) that was identified in the gas phase products of previous $N_2$-$CH_4$-CO experiments (Bernard et al. 2003; Coll et al. 2003). The intensity ratio of the peaks at 44, 45, and 46 amu is 100: (1.16): (0.36), which matches to the isotopic distribution of $CO_2$ (100: 1.2: 0.4). Thus, the peak at 44 amu is contributed mainly by $CO_2$ in our experiments, and the peaks at 45 and 46 amu are the isotopic peaks of $CO_2$. The peak at 1 amu is from hydrogen (H), a fragment of $H_2$ (2 amu), produced by electron-impact in the RGA mass spectrometer. Similarly, the peak at 12 amu is from carbon (C), a fragment of CO and/or $CH_4$. The intensity change of these peaks (1, 12, 45 and 46 amu) is caused by the change of $H_2$, CO and $CO_2$ in the gas phase. The amount of $H_2$ decreases while $H_2O$, HCN, $C_2H_5N$/HCNO and $CO_2$ increase in the gas phase, when CO increases from 0 to 5% (Figure 2A). Their intensity change with CO concentration is plotted in Figure 2B, showing that the peak intensity changes gradually with the increase of CO. As CO increases in the initial gases, less $H_2$ but more $H_2O$, HCN, $C_2H_5N$/HCNO and $CO_2$ are produced. The changes in the gas phase affect the formation and growth of the solid particles.

*3.2. Production Rate, Density and Elemental Compositions of the Solids*

From the mass of the solid sample we collected and the experiment running time (~72 hours), we calculated the production rate of the solids in seven experiments, which varies between 7.25 and 7.42 mg h$^{-1}$. The differences in production rate are smaller than the error (0.20 mg h$^{-1}$) induced from the deviation of the balance and the solid sample residuals on the wall of the reaction chamber; the production rate does not appear to change significantly with the introduction of CO. However, the similarity in production rate does not mean the solid samples from the seven experiments are the same. For example, their density is different.

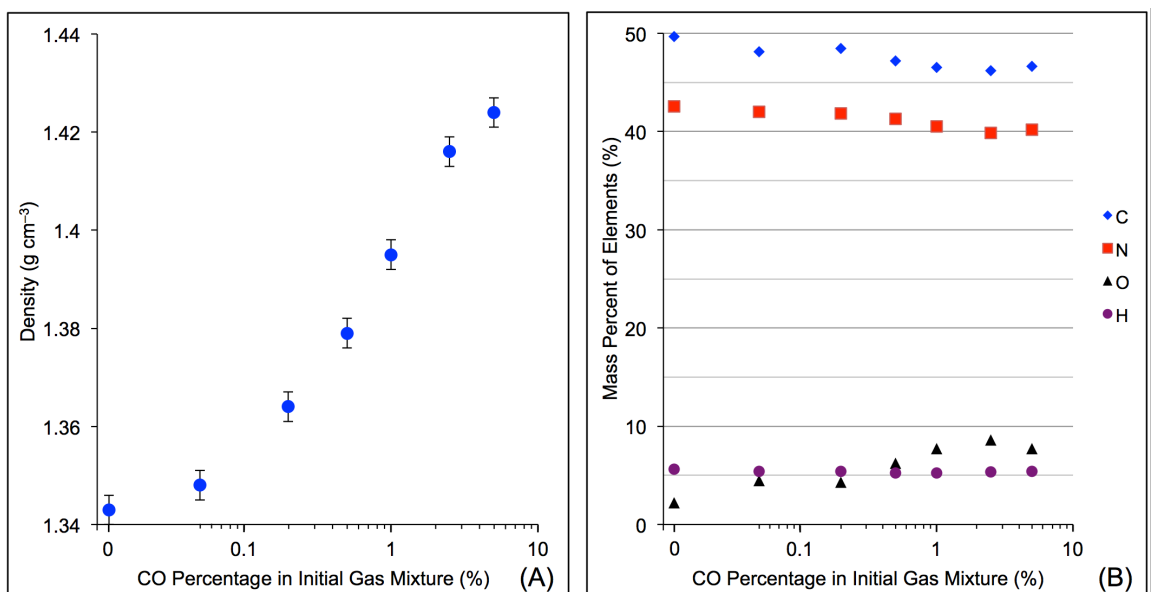

**Figure 3.** (A) The solid density variation with the increase of CO in the initial gases; (B) The mass percentages of C, H, N and O in the solid samples change with the increase of CO in the initial gases (Error bars are 0.5% for elemental compositions from replicate runs, and are the same size or smaller than the symbols).

As shown in Figure 3A, the density of the solid sample increases about 6%, from 1.343 g cm$^{-3}$ (no CO) to 1.424 g cm$^{-3}$ (5% CO). According to the definition of density (DeCarlo et al. 2004; Hörst and Tolbert 2013), the density we measured here is either particle density (if particles have internal voids) or material density (if particles have no internal voids). Imanaka et al. (2012) measured the density of tholin (produced from 10% CH$_4$ in N$_2$ at pressures of 1.6 and 23 mbar) using a gas pycnometer and found densities of 1.31–1.38 g cm$^{-3}$. The density of our solid sample (produced from 5% CH$_4$ in N$_2$, 1.343 g cm$^{-3}$) falls in that range. Using a different technique (aerosol mass spectrometer and scanning mobility particle sizer), Trainer et al. (2006) and Hörst and Tolbert (2013) found the effective particle density of tholin particles (produced from CH$_4$ in N$_2$) is in the range from 0.5 to 1.1 g cm$^{-3}$, which is lower than that we measured. This could be caused by different setup/conditions used for producing tholins, and by different methods of density measurement. Additionally, the solid materials we used for density measurements are collected from tholin films deposited on the chamber wall. The particle deposition process may reduce internal voids and porosity and thus increase the density. Our measurement shows that the density increases gradually with the increase of CO. Figure 3B shows the mass percent of C, H, N and O in seven solid samples. As CO increases, the carbon and nitrogen content in the solid samples decreases while the oxygen content increases, but the hydrogen content remains the same at ~5%. Oxygen and water contamination are well known problems for tholin production and analysis (Cable et al. 2012). Even though we try to minimize the oxygen and water contamination in the process of sample production and collection, there is still about 2% of O in the solids produced without CO in the initial gases. The 2% oxygen contamination in the sample is mostly like from the water adsorbed during sample analysis, and is a smaller degree of contamination than other studies (Sciamma-O'Brien et al. 2010; Fleury et al. 2014). As CO increases, the oxygen content in the solid sample increases from 2% to ~8%, indicating a significant oxygen incorporation in the solids. As CO increases, the elemental analysis results show a 2% decrease of N-content in the solids, while MS results show an increase of two possible N-containing gas products. It is likely that with the inclusion of CO, there are O-species competing with N-species (radicals and ions) in the system, leading to less nitrogen incorporation in the solid and generating more N-containing products in the gases. When CO increases in the initial gases, more O atoms replace C or N atoms in the solids. Since oxygen is the heaviest element among C, H, N and O, the elemental change in the solids could increase the density by 1.9% at most. This is smaller than the 6% density increase, which indicates that there are other factors affecting the density, such as molecular weight, polarity, and degree of unsaturation.

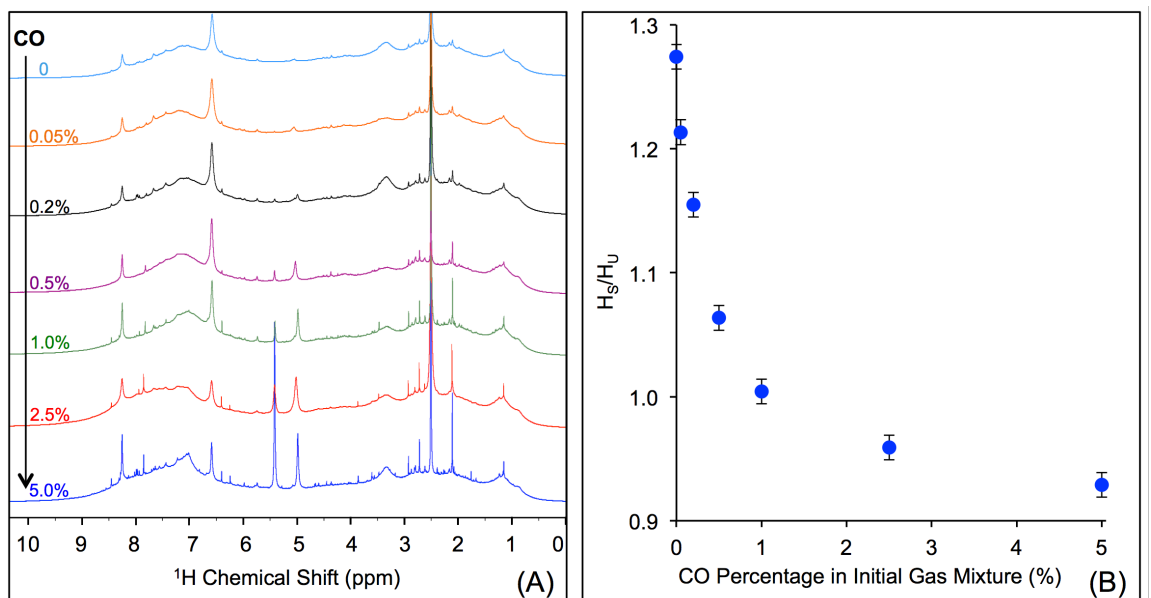

**Figure 4.** (A) 1D $^1$H NMR spectra of the solids in DMSO-$d_6$ (Stacked from top to down, CO concentration in the initial gases increases from 0 to 5%); (B) The ratio of the hydrogen atoms in saturated structures to the hydrogen atoms in unsaturated structures ($H_S/H_U$) in the solids change with the increase of CO in the initial gases.

*3.3. NMR Result of the solids*

Figure 4A shows the 1D $^1$H NMR spectra of the solids dissolved in DMSO-$d_6$. The signals, including sharp peaks (from small molecules) and broad peaks (from larger polymeric structures), are distributed between 0.5 and 9 ppm, which indicates that there are various and complex structures in the samples (He & Smith 2014a). The chemical shift in NMR spectroscopy indicates the local chemical bonding environment of the nuclei. Specifically, the $^1$H chemical shift in our NMR spectra reveals the hybridization status of carbon or nitrogen, which is bonded to the hydrogen directly and/or 2 bonds away from the hydrogen (Lambert & Mazzola 2004). Based on this, we separate broad and sharp peaks into a saturated region (0–5 ppm) and an unsaturated region (5–9 ppm). The signals at 0.5–5 ppm are usually derived from protons in saturated structures, such as saturated hydrocarbons, saturated amines and alcohols, while the signals above 5 ppm are from protons in unsaturated structures such as aromatics, imines, and $NH_2$ or OH bonded to unsaturated carbon (He et al., 2012; Lambert & Mazzola 2004). The number and intensity of sharp peaks increase from top to bottom in Figure 4A, indicating that the variety and amount of small molecules increase with the increase of CO. Some sharp peaks increase significantly, such as those at 2.11, 5.01, 5.41, and 8.27 ppm. The peaks at 5.41 and 8.27 ppm could be derived respectively from guanidine hydrocyanide ($C_2H_6N_4$) and 1,2,4-triazole ($C_2H_3N_3$), as reported in previous NMR study of tholin (He and Smith 2013, 2014b). Both molecules are heavily nitrogenated unsaturated compounds, showing that the unsaturated species are clearly increasing with the increase of CO. Since the peaks in the spectra are quantitative, we integrate the total peak areas for saturated and unsaturated regions in each spectrum and get a ratio of the hydrogen atoms in saturated structures to those in unsaturated structures ($H_S/H_U$) in the solids. As shown in Figure 4B, the $H_S/H_U$ ratio decreases as CO increases. Because the total hydrogen content in the samples stays the same (as demonstrated by elemental analysis, see 3.2), the decreasing $H_S/H_U$ ratio demonstrates that the hydrogen atoms in saturated structures decrease while those in unsaturated structures increase with the increase of CO, indicating that the inclusion of CO in the initial gases increases the degree of the unsaturation in the solids.

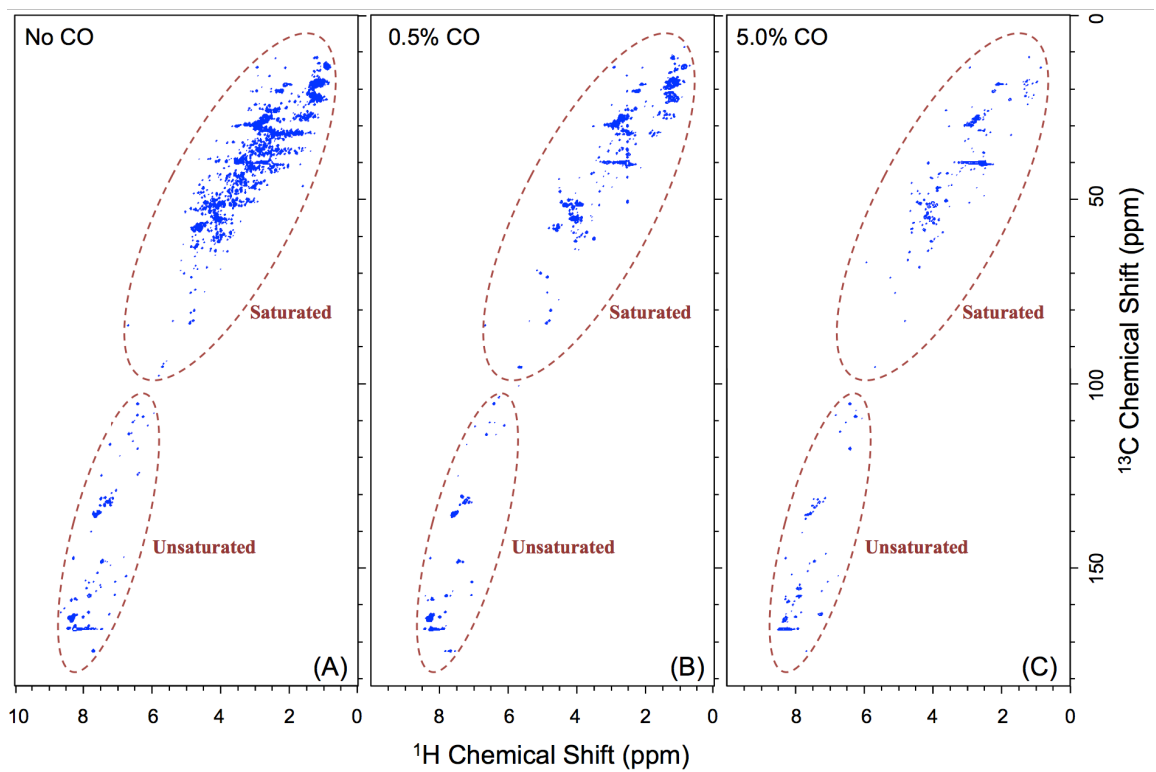

**Figure 5.** 2D $^1$H–$^{13}$C HSQC NMR spectra of three solids samples with different CO concentrations in the initial gases: (A) no CO, (B) 0.5%, and (C) 5%. Each signal represents a proton that is bound to a carbon atom. Based on the chemical shift of both proton and carbon, the signals on the spectra are divided into two groups: saturated ($^1$H: 0–5 ppm, $^{13}$C: 10–90 ppm) and unsaturated ($^1$H: 5–9 ppm, $^{13}$C: 100–180 ppm) structures (Breitmaier and Voelter 1987; Lambert & Mazzola 2004). The signals in the saturated groups are from hydrogen bonded to saturated carbons (sp$^3$-hybridized) while the signals in the unsaturated groups are from the hydrogen bonded to unsaturated carbons (sp$^2$- and sp-hybridized).

Figure 5 shows the 2D $^1$H–$^{13}$C HSQC NMR spectra of three solids samples with different CO concentrations, no CO, 0.5%, and 5%, respectively. The $^1$H signal distribution is consistent with that in the $^1$H NMR spectra. The $^{13}$C signals are distributed between 10 and 180 ppm, which confirms the variety and complexity of the structures in the samples (Breitmaier and Voelter 1987). The carbon signals at 10–100 ppm are usually derived from saturated carbon (sp$^3$-hybridized) while the signals above 100 ppm are from unsaturated carbon (sp$^2$ or sp-hybridized). According to the chemical shift of both proton and carbon (Breitmaier and Voelter 1987; Lambert & Mazzola 2004), the signals on the spectra are divided into saturated and unsaturated groups. Figure 5 indicates that the hydrogen atoms bonded to saturated carbons decrease significantly while those bonded to unsaturated carbons do not change much as CO increases. We already learned from the $^1$H NMR spectra (Figure 4) that the hydrogen atoms in saturated structures decrease while those in unsaturated structures increase as CO increases. The results from Figure 5 illustrate that the decrease of the hydrogen atoms in saturated structures is due to fewer hydrogen atoms bonded to saturated carbons, but the hydrogen atoms bonded to unsaturated carbons do not cause the increase of the hydrogen atoms in unsaturated structures. Since there are only H, C, N and O in the samples, the increase of the hydrogen atoms in unsaturated structures can only be induced by more hydrogen atoms bonded to nitrogen or oxygen in unsaturated structures, such as aromatic amines, imines, and phenols (Lambert & Mazzola 2004).

*3.4 CO Affecting Planetary Atmospheric Chemistry*

It is reported that the initial amount of CO affects the production rate of tholins. Hörst and Tolbert (2013) found that the inclusion of CO increases both particle size and number density, while Fleury et al. (2014) reported that the presence of CO drastically decreases the production of tholins. However, our experiments indicate that the concentration of CO does not affect the production rate of the solids. To be noted, experiments in three studies were performed on different setups and at different conditions, and production rate could not provide details on the chemistry leading to the formation of the solid particles. Our results demonstrate that the inclusion of CO affects the chemistry in both the gas phase and the solid phase. With the increase of CO in the initial gases, there is less $H_2$ but more $H_2O$, HCN, $C_2H_5N$/HCNO and $CO_2$ produced in the gas phase, while the density, oxygen content and degree of unsaturation of the solids increase. These results reveal how CO could affect the chemical processes in the system.

Under electrical discharge, CO is dissociated to carbon and oxygen radicals. Carbon radicals tend to form triple-bond and double-bond with nitrogen, leading to the production of HCN and $C_2H_5N$/HCNO, and more unsaturated large molecules in the solids. This fact leads to the increasing unsaturation of the solids, as the NMR results indicated. Oxygen radicals react with $H_2$, removing $H_2$ from the system. Decreasing $H_2$ make the system less reducing, which could also contribute to the increased unsaturation of the solids. After reactions with other species, oxygen radicals end up in $H_2O$, $CO_2$, and contribute to further oxygen incorporation in the solids. The oxygen bearing species in the system could terminate molecules and inhibit the growth of long-chains (Trainer et al., 2006; Fleury et al., 2014), resulting in more small molecules in the solids as observed in 1D $^1H$ NMR spectra (Figure 4A). The discharge used in our experiments could simulate energy sources in atmospheres, such as charged particles from the sun and the universe, magnetospheric charged particles, and lightning. The discharge here can break the triple-bond in $N_2$ and CO to simulate upper atmosphere chemistry, while far ultraviolet lamps (115-400 nm) used in the laboratory have difficulties to break these bonds (Cable et al., 2012; Hörst & Tolbert 2013, 2014). Therefore, the processes happening in our experiments could also occur in planetary atmospheres and produce organic molecules containing oxygen, including potential prebiotic molecules (Hörst et al., 2012).

4. CONCLUSIONS

We investigated the effect of CO on both gas and solid phase chemistry in a series of planetary atmosphere simulation experiments using gas mixtures of CO, $CH_4$, and $N_2$ with a range of CO mixing ratios from 0.05% to 5% at low temperature (~100 K). Our results show that the inclusion of CO has a dramatic effect on the gas phase chemistry, as well as the density and composition of the solids. Specifically, with the increase of CO in the initial gases, there is less $H_2$ but more $H_2O$, HCN, $C_2H_5N$/HCNO and $CO_2$ produced in the gas phase, while the density, oxygen content and degree of unsaturation of the solids increase. CO offers an extra carbon source and creates a less reducing environment, allowing more unsaturated species to be produced. CO also serves as an oxygen source for oxygen incorporation in both gas and solid phases, and the produced oxygen bearing species terminate molecules and inhibit the growth of long-chains, yielding more small molecules in the solids. The results here provide insight into the effect of CO on the chemical processes taking place in our experiments, and these processes could have occurred or be occurring in planetary atmospheres.

C. H. is supported by the Morton K. and Jane Blaustein Foundation. J. S. and N. P. acknowledge support from NASA through the Iowa Space Grant Consortium (Grant No. NNX16AL88H).